\documentclass[doublecol]{epl2}
\usepackage{graphicx}
\usepackage{amssymb}
\usepackage{amsmath}

\newcommand{\be}{\begin{equation}}
\newcommand{\ee}{\end{equation}}
\newcommand{\bea}{\begin{eqnarray}}
\newcommand{\eea}{\end{eqnarray}}
\newcommand{\besa}{\begin{subeqnarray}}
\newcommand{\eesa}{\end{subeqnarray}}
\newcommand{\bean}{\begin{eqnarray*}}
\newcommand{\eean}{\end{eqnarray*}}

\title{Exact equivalence between one-dimensional Bose gases interacting via hard-sphere and zero-range potentials}
\shorttitle{Exact equivalence between one-dimensional hard-sphere and zero-range potentials}

\author{Manuel Valiente}

\shortauthor{M. Valiente}

\institute{Lundbeck Foundation Theoretical Center for Quantum System Research, Department of Physics and Astronomy, Aarhus University, DK-8000 Aarhus C, Denmark, EU}


\abstract{
We prove the equivalence between the hard-sphere Bose gas and a system with momentum-dependent zero-range interactions in one spatial dimension, which we call extended hard-sphere Bose gas. The two-body interaction in the latter model has the advantage of being a regular pseudopotential. The most immediate consequence is the existence of its Fourier transform, permitting the formulation of the problem in momentum space, not possible with the original hard-core interaction. In addition, in the extended system, interactions are defined in terms of the scattering length, positive or negative, identified with the hard-sphere diameter only when it is positive. We are then able to obtain, directly in the thermodynamic limit, the ground state energy of the strongly repulsive Lieb-Liniger gas and, more importantly, the energy of the lowest-lying super Tonks-Girardeau gas state with finite, strongly attractive interactions, in perturbation theory from the novel extended hard-sphere Bose gas.     
}

\begin{document}
\maketitle

\section{Introduction} 
Low-dimensional systems have been very popular among theorists for a number of decades now \cite{Mattis,Cazalilla}, not without a reason. On one hand, many of these allow for an exact treatment. The Bethe {\it ansatz} solves a number of one-dimensional problems exactly, such as the Lieb-Liniger gas \cite{LiebLiniger} or Heisenberg's model \cite{IntroBethe}, while systems with supersymmetric Hamiltonians \cite{Cooper}, such as Sutherland's model \cite{Sutherlandpaper} or the attractive Lieb-Liniger model \cite{Mattisprivate}, allow for a trivial evaluation of their ground state wave functions \cite{Mattis}. On the other hand, one-dimensional systems have physical properties of interest, and can become strongly correlated, as is the case of the Tonks-Girardeau gas \cite{Girardeau,Yukalov} -- a one-dimensional system of impenetrable bosons sharing many of its properties with the ideal Fermi gas -- which was experimentally realized in \cite{Paredes}. More recently, the metastability of the so-called super Tonks-Girardeau gas -- a strongly attractive one-dimensional Bose gas with no bound clusters -- has been proposed \cite{Astra1,Batchelor,Astra2} and subsequently experimentally realized \cite{Haller}. This unique system attracts much of current interest \cite{Chen,Kormos,Yin,Carnicero}.

There is strong theoretical evidence \cite{Batchelor} on the close relation between the lowest-lying super Tonks-Girardeau (sTG) gas state and a system of one-dimensional (1D) hard-spheres in the low-density limit. This is indeed very appealing, since the hard-sphere gas is extremely simple to solve with a variety of methods (see e.g. \cite{Girardeau}), while a good description of the 1D Bose gas in the sTG regime requires, in principle, accurate numerical calculations, such as the diffusion and variational Monte Carlo employed in \cite{Astra2}. This evidence was then used to conjecture that, in a harmonic trap, the sTG and 1D hard-sphere Bose gases are equivalent \cite{AstraGirardeau}; we call this statement Astrakharchik-Girardeau conjecture for the trapped case.

In this Letter, we prove a theorem which states the equivalence between 1D Bose gases interacting via hard-sphere potentials and certain momentum-dependent contact interactions -- which we also define in general -- in the untrapped case. Our theorem represents a weaker, though exact version of the Astrakharchik-Girardeau conjecture in free space. The two-body interactions in our new model, which we call extended hard-sphere (eHS) system, are Fourier-transformable and we are therefore able to write down the system's Hamiltonian in momentum representation. For the Lieb-Liniger gas, we obtain the energy of both the ground state with strong repulsive interactions and the lowest sTG state with strong attraction to first-order in perturbation theory, departing from the exactly-solvable eHS model as the zero-th order reference Hamiltonian. Last but not least, we obtain the so-called Tan relations \cite{Olsha2,Tan1,Tan2,Tan3,Braaten,Castin,Combescot,VZM,Zwerger} for the Lieb-Liniger gas and use them to calculate Tan's contact -- related to short-distance correlations -- in the homogeneous and trapped cases, the latter within the local density approximation.       

\section{System Hamiltonians}
The Hamiltonians for the hard-sphere (HS) and Lieb-Liniger (LL) systems with $N$ identical bosons of mass $m$ have the form
\be
H=\sum_{i=1}^N \frac{\hat{p}_i^2}{2m}+\sum_{i<j=1}^N V(x_i-x_j).\label{Ham1}
\ee
For the LL model, $V(x)= g_{\text{LL}} \delta(x)$, $g_{\text{LL}}=-2\hbar^2/ma$ constant, and $a$ the two-body scattering length; for the HS model, $V(x)=0$ ($\infty$) for $|x|>a$ ($\le a$), with $a>0$ a constant hard-sphere diameter.
 
\section{Two-body problem} We begin by considering the two-boson problem. After separation of center-of-mass ($X=(x_1+x_2)/2$) and relative ($x=x_1-x_2$) coordinates, Hamiltonian (\ref{Ham1}) reads
\be
H=-\frac{\hbar^2}{2\mu} \frac{\partial^2}{\partial x^2}+V(x),\label{HamTwoBody}
\ee
with $\mu=m/2$ the reduced mass of the two-boson system. The stationary Schr\"odinger equation $H\psi=E\psi$ at positive energies $E=\hbar^2k^2/2\mu$ is solved by $\psi(x)=\sin(k|x|+\theta_{\alpha})$. The wave functions of the HS model are given by $\psi(x)$ for $|x|>a$ and are zero for $|x|\le a$, while for the LL model they are given by $\psi(x)$ for all $x$. The phase shifts, with self-explanatory subscripts, are given by
\begin{align}
\tan\theta_{\text{LL}} &= \frac{\hbar^2 k}{\mu g_{\text{LL}}} \label{phaseshiftsLL}\\
\tan\theta_{\text{HS}} &= -\tan(ka).\label{phaseshiftsHS}
\end{align}

Note that if, in Eq. (\ref{phaseshiftsLL}), $g_{\text{LL}}$ is made momentum-dependent, $g_{\text{LL}}\to g(k)$, then the HS phase-shifts (\ref{phaseshiftsHS}) are recovered by choosing
\be
g(k)=-\frac{\hbar^2k}{\mu}\cot(ka).\label{momentumdependentg}
\ee
The above relation, carefully stated, provides the desired mapping between HS and Dirac delta interactions in one dimension.

\section{Definition of momentum-dependent interactions} 
Any analytic function of an operator is defined by its expansion in powers of the operator \cite{Conway}.
Given that, define an even analytic function of the momentum operator $f(\hat{k})$, with $\hat{k}=\hat{p}/\hbar$. The action of $f(\hat{k})$ on a plane-wave is given by $f(\hat{k})e^{iqx}=f(q) e^{iqx}$.
It is also necessary to consider its action on functions of the form $\sin(q|x|)$, with discontinuous derivatives at the origin, which is given by $\sum_{n=0}^{\infty}(-1)^nf^{(2n)}(0) (\partial_{x})^{2n} \sin(q|x|)/(2n)!$.
All even derivatives of $\sin(q|x|)$ include undesired Dirac deltas $\delta(x)$. A simple way to deal with this problem is to restrict the action of $f$ to positions $x>0$ or $x<0$, which avoids Dirac deltas and has no influence on differentiable functions ($\sim \cos(kx)$) at the origin. We can therefore define momentum-dependent contact interactions as follows:  

{\it Definition}. Let $f$ be an even, analytic function. A momentum-dependent contact interaction is an operator $\hat{W}_k$ with the action
\be
[\hat{W}_k\psi](x) = \delta(x) \lim_{x\to 0^+} f\left(-i\frac{\partial}{\partial x}\right) \psi(x).\label{momentumdependentW}
\ee

With the above definition, $\hat{W}_k$ is not Hermitian. This is a general property of momentum-dependent pseudopotentials, shared by the famous partial-wave pseudopotentials of Huang and Yang \cite{HuangYang}.
  
\section{Equivalence between hard-sphere and contact interactions} From the above considerations, the proof of the equivalence between the HS states and the states corresponding to momentum-dependent contact interactions for the two-body case is immediate:

{\it Theorem}. Let $H$ be defined by Eq. (\ref{HamTwoBody}), with 
\be
V(x)=\delta(x) g(\hat{k}),\label{potential1}
\ee
where $g(\hat{k})=-\hbar^2 \hat{k} \cot{a\hat{k}}$, and the limit $x\to 0^+$ as in Eq. (\ref{momentumdependentW}) is assumed. Let $\psi_k:\mathbb{R}\to \mathbb{C}$ be the bosonic scattering wave functions of $H$, at energies $\hbar^2k^2/2\mu$, $k$ real. Then, the restrictions $\phi_k$ of $\psi_k$ to $\mathcal{D}=\mathbb{R}-(-a,a)$, $a>0$, are the bosonic scattering wave functions at the same energies for the problem of two hard spheres of diameter $a$.

Remarks: (i) $a>0$ is necessary for the mapping, although Hamiltonian (\ref{HamTwoBody}) with potential (\ref{potential1}) is well-defined for $a<0$, so we regard $a$ as the scattering length of the model, which we call extended hard-sphere (eHS) model; (ii) it is fundamental in the above theorem that the energies are positive, since there exists an unphysical, infinitely-bound state which becomes the identically zero function in the restricted domain of the HS wave-functions.  

\section{Many-body problem}
Fortunately, the many-boson problem with contact interactions is exactly solvable by
means of the Bethe {\it ansatz}. This leaves us with the equivalence of the HS and the extended model, stated as follows 

{\it Theorem}. Let $H$ be defined by Eq. (\ref{Ham1}), with
\be
V(x_i-x_j) = \delta(x_i-x_j) g(\hat{k}_{ij}),\label{potential2}
\ee
where $k_{ij}=(k_i-k_j)/2$, $g(\hat{k})=-\hbar^2 \hat{k} \cot a\hat{k}$, and the limit $x_i-x_j\to 0^+$ as in Eq. (\ref{momentumdependentW}) is assumed. Let
$\psi_{k_1,k_2,\ldots,k_N}:\mathbb{R}^N \to \mathbb{C}$ be a bosonic scattering wave
function of $H$ with energy $E=\sum_{i=1}^N \hbar^2 k_i^2 / 2m$ and $k_i$ real,
$i=1,\ldots,N$. Then, the restrictions $\phi_{k_1,k_2,\ldots,k_N}$ of
$\psi_{k_1,k_2,\ldots,k_N}$ to the domain $\mathcal{D}=\mathbb{R}^N-\cup_{i<j=1}^N
\mathcal{I}^{a}_{i,j}$, with 
\be
\mathcal{I}^{a}_{i,j}=\{(x_1,x_2,\ldots,x_N) \in \mathbb{R}^N | |x_i-x_j|<a \}, 
\ee
are the bosonic scattering states for the $N$-boson problem with HS interactions of
diameter $a>0$ at the same energies. The same conclusion holds valid in a box of length $L>Na$.

{\it Proof}. Showing that Bethe {\it ansatz} (BA) wave functions constitute the bosonic scattering states
of Hamiltonian (\ref{Ham1}) with interactions (\ref{potential2}) is trivial: the momentum-dependent interactions only
multiply each term of the BA by a constant (which depends on the relative momenta,
obviously). Since the potentials have a zero range, scattering occurs without diffraction \cite{BeautifulModels} and therefore the model is exactly solvable via BA. For the
finite box case, the same holds evidently true. Now, because the two-body scattering
phase shifts for the eHS model are identical to the HS phase shifts, the BA
equations are identical, too. 
In the HS case, the saturation density, for which the ground state energy at finite
densities (in a finite box) diverges, is given by $N/L = 1/a$. At that point the
wave functions simply do not exist.
Now assume that $N/L<a$ and take the restriction $\phi_{k_1,\ldots,k_N}$ of a BA wave
function $\psi_{k_1,\ldots,k_N}$ to $\mathcal{D}$. Since the BA equations for the
two models are identical, $\phi_{k_1,\ldots,k_N}$ are the scattering wave functions for
the HS problem. QED. 

\section{Momentum space} We now turn to Fourier-transform any general momentum-dependent contact interactions, including the particular eHS interactions. Our starting point is the interaction in terms of the bosonic field operators $\hat{\psi}$,
\be
W_k=\frac{1}{2} \int \hat{\psi}^{\dagger}(x) \hat{\psi}^{\dagger}(x') W_k(x-x')  \hat{\psi}(x) \hat{\psi}(x') dx' dx,\label{potentialspace}
\ee
The field operators are expanded in the plane-wave basis as $\hat{\psi}(x) = \frac{1}{\sqrt{L}} \sum_{p} a_p e^{i p x}$, with $L$ the length of the system, and $a_p$ the bosonic annihilation operators in momentum space. Inserting this expansion into Eq. (\ref{potentialspace}), we obtain the momentum representation of $W_k$,
\be
W_k = \frac{1}{2L} \sum_{p_1,p_2,q} g(p_{1,2}) a_{p_1+q}^{\dagger}a_{p_2-q}^{\dagger}a_{p_1}a_{p_2},\label{fourier}
\ee
where $p_{1,2}\equiv (p_1-p_2)/2$. In particular, Eq. (\ref{fourier}) applied to the extended HS interactions provides, for the first time, a simple momentum representation for a very singular potential.

\section{Strongly-interacting delta Bose gas as a perturbation from the hard-sphere model}
A major inconvenience of using singular interactions such as the HS potential is that it is not possible to perform perturbation expansions in the small parameter $\rho a$, with $\rho=N/L$ the particle density. Remarkably, the ground-state energy of the strongly repulsive ($-\rho a \ll 1$) LL gas coincides asymptotically with the ground state energy of the HS model, albeit with $a<0$ \cite{LiebLiniger}, that is, the eHS model with $a<0$. With the novel momentum-dependent interactions developed in this work, we are able to perform perturbation theory for the strongly attractive or repulsive LL gas, starting from the exact ground state for the eHS model, for both $a>0$ and $a<0$. The success of such a perturbative treatment, which is seen {\it a posteriori}, is {\it a priori} expected, since the maximal difference of two-body T-matrices for the eHS and LL models at density $\rho$ behaves as $(T_{\mathrm{eHS}}-T_{\mathrm{LL}})/\rho = O[(\rho a)^3]$, while the maximal difference between fermionized (Tonks-Girardeau) and LL T-matrices behaves as $(T_{\mathrm{TG}}-T_{\mathrm{LL}})/\rho = O(\rho a)$, which is two orders worse in the small $\rho a$. These estimates show that, effectively, first-order perturbation theory from the eHS model corresponds to a third-order expansion from the fermionized Bose gas. 

The LL Hamiltonian $H_{\text{LL}}$ can be written in terms of the eHS Hamiltonian $H_{\text{eHS}}$ (Eq. (\ref{Ham1}) with momentum-dependent interactions (\ref{momentumdependentg})) as
\be
H_{\text{LL}} = H_{\text{eHS}}+\sum_{i<j=1}^N\delta(x_i-x_j) [g_{\text{LL}}-g(\hat{k}_{ij})],
\ee
where $g_{\text{LL}}$ is the LL interaction strength and $g(\hat{k}_{ij})$ is given by Eq. (\ref{momentumdependentg}), with the limit $x_i-x_j\to 0^+$ as in Eq. (\ref{momentumdependentW}) implicitly assumed. Making use of the bosonic symmetry of the particles, we can show that in any state of the eHS Hamiltonian, the expectation value $E^{(1)}$ of the perturbation is given by $E^{(1)}=\langle{\delta(x_1-x_2)\rangle} \sum_{i<j=1}^N\tilde{g}(k_{ij})$, with $\{k_{ij}\}_{ij}$ the set of BA relative momenta of the particular eHS (or HS) state, and $\tilde{g}(k_{ij})=g_{\text{LL}}-g(k_{ij})$. In the thermodynamic limit (TL) $E^{(1)}$ is well-defined for $\rho a <1/2$, which sets an upper bound on the radius of convergence of the perturbation expansion. We now apply the Hellmann-Feynman theorem, $dE_{\text{HS}}/da = \sum_{i<j}\langle \delta(x_i-x_j)\rangle dg(\hat{k}_{ij})/da$, where $E_{\text{HS}}$ is the ground state energy of the HS Bose gas, given by \cite{Girardeau}
\begin{equation}
E_{\text{HS}}=\frac{\pi^2\hbar^2\rho^2}{6m(1-\rho a)^2}\frac{N^2-1}{N}.\label{energyHS}
\end{equation}
For the first-order energy correction we obtain $E^{(1)}= \mathcal{C}(\rho,a) dE_{\text{HS}}/da$, with
\be
\mathcal{C}(\rho,a) = \frac{\sum_{i<j=1}^{N} \tilde{g}(k_{ij})}{\sum_{i<j=1}^N \frac{dg}{d a}(k_{ij})}.\label{Ca}
\ee
In the TL, the density of states (in the BA \cite{BeautifulModels}) for the HS Bose gas is a constant times a step function, and therefore we can write
\be
\mathcal{C}(\rho,a) = \frac{\int_{-q}^{q}dk_1 \int_{-q}^{q} dk_2 \tilde{g}(k_{12})}{\int_{-q}^{q} dk_1 \int_{-q}^{q} dk_2 \frac{dg}{d a}(k_{12})}\approx -\frac{a(qa)^2}{18+(qa)^2},\label{Caint}
\ee
where $q=\pi \rho /(1-a\rho)$, and where the approximate equality is valid for $q|a|\ll 1$.  

\begin{figure}
\includegraphics[width=0.44\textwidth]{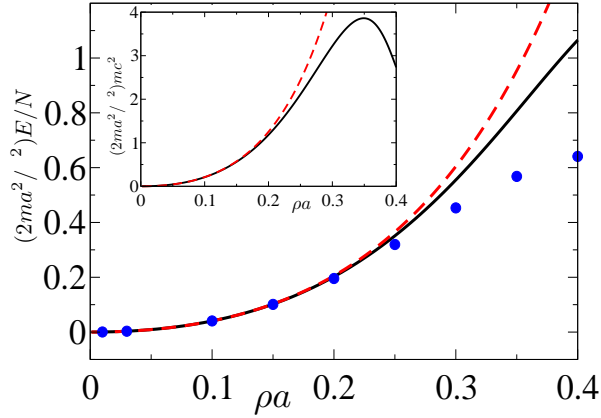}
\caption{Main figure: energy per particle as a function of $\rho a$ for sTG gas in 1st order perturbation theory (black solid line), compared to HS result (dashed red line) and Monte Carlo results of ref. \cite{Astra2} (blue dots). Inset: inverse compresibility for sTG to 1st order (black solid line) and for HS gas (red dashed line).}
\label{fig}
\end{figure}

The perturbative correction for the repulsive LL gas ($a<0$) yields a minor improvement with respect to the eHS asymptotic value, so we concentrate on the attractive case, i.e. the sTG gas. In Fig. \ref{fig} we show the energy per particle in the TL to first order in perturbation theory, $E/N\approx (E_{\text{HS}} + E^{(1)})/N$, as a function of the gas parameter $\rho a$, and compare our results with existing Variational Monte Carlo (VMC) data from ref. \cite{Astra2} and with the HS result, Eq. (\ref{energyHS}). Our results are in excellent agreement with the VMC calculations up to $\rho a \approx 0.25$. Beyond that point and until $\rho a \approx 0.3$ our results deviate from VMC, but their tendency is still correct. For $\rho a > 0.3$ our calculation is not enough to describe the sTG gas, since in the present case the energy is overestimated in this region. In Fig. \ref{fig} we also show the inverse compressibility $mc^2=\rho \partial_{\rho} \mu$, with $\mu$ the chemical potential, as a function of $\rho a$ to first order in perturbation theory. Our results reproduce the overall features calculated in \cite{Astra2} and are in good agreement until $\rho a \approx 0.2$, from where our calculation largely overestimates the fitted VMC results.

\section{Tan's contact for the sTG gas}\label{sectionTan}
A quantity which has attracted much of recent theoretical and experimental interest is the so-called contact \cite{Tan1}, denoted by $\mathcal{I}$. This is defined for bosonic and fermionic systems with zero-range interactions in any dimension \cite{Olsha2,Tan1,Castin,Combescot,VZM,Zwerger} as the coefficient of the asymptotic part of the momentum distribution $n_{\sigma}(\mathbf{k})$, $\mathcal{I}=\Omega\lim_{|\mathbf{k}|\to \infty} k^4 n_{\sigma}(\mathbf{k})$,
where $\Omega=L^D$ is the volume, $D$ is the dimension and $\sigma$ is the spin component (omitted for spinless bosons). In the 1D case, relevant here, it is related to short-distance correlations \cite{Olsha2,Gora}. Relations between different properties of the system and the contact are generally known as Tan relations. We focus here in Tan relations for the LL gas with or without a trap, which we then apply to the sTG gas.

Tan relations can be proved in parallel to the higher-dimensional cases \cite{Tan1,Tan2,Tan3,VZM} by using the 1D version of the so-called $\eta$-selector \cite{Tan1,ValienteTan,VZM,Tanarxiv}. This reads 
\be
\eta(k)=1+\frac{\pi \hbar}{mg_{\mathrm{LL}}}\delta(1/|k|).
\ee
The energy of the system is given by 
\be
E=\frac{\hbar^2}{2m}\sum_k \eta(k)k^2 n_k + \langle \mathcal{W} \rangle,
\ee
with $\mathcal{W}\equiv \sum_{i=1}^N W(x_i)$ the total single-particle trapping potential. The adiabatic energy theorem reads
\be
\frac{dE}{da}=\frac{\hbar^2 \mathcal{I}}{m}. 
\ee
The two relations above were already known for the homogeneous case \cite{LiebLiniger,Olsha2}. The generalized virial theorem, assuming $W(x)\propto x^{\beta}$, is given by
\be
E=\frac{\beta + 2}{2} \langle \mathcal{W} \rangle -\frac{\hbar^2\mathcal{I}}{2m} a.\label{virial}
\ee
Last, the pressure relation, which is only valid in the homogeneous case, is given by 
\be
PL=2E+\frac{\hbar^2\mathcal{I}a}{2m}.
\ee

\begin{figure}
\includegraphics[width=0.44\textwidth]{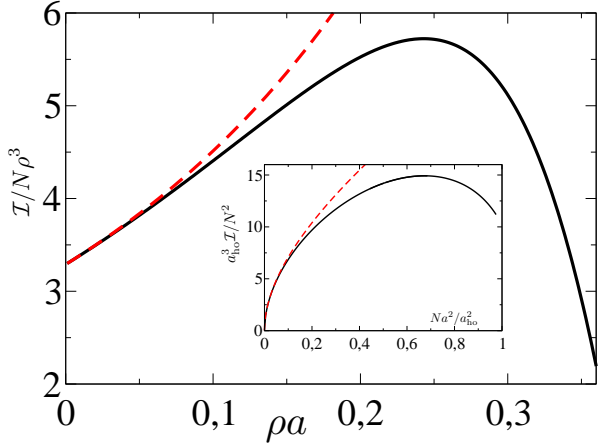}
\caption{Main figure: contact per particle as a function of $\rho a$ for sTG gas in 1st order perturbation theory (black solid line), compared to HS result (dashed red line). Inset: contact for harmonically trapped sTG to 1st order in the LDA from virial theorem, Eq. (\ref{virial}), (black solid line) compared to contact for trapped HS gas (red dashed line).}
\label{fig-contact}
\end{figure}

In Fig. \ref{fig-contact}, we plot the contact per particle $\mathcal{I}/N$ for the homogeneous sTG gas obtained via perturbation theory from the eHS gas in the thermodynamic limit. As expected, deviations of the perturbative contact from the asymptotic HS result are more pronounced than for the energy as the gas parameter grows. The contact for a given quantum state of a system and, in particular, for a stationary state of the Schr\"odinger equation, is a positive quantity. A violation of this property implies that the state under consideration is not a physical state of the system. As observed in Fig. \ref{fig-contact}, the contact exhibits a maximum at $\rho a\approx 0.25$, where perturbation theory is still approximately correct, showing that it will eventually become negative at a given critical value of the gas parameter $(\rho a)_c$ where a super-Tonks-Girardeau gas cannot exist. The VMC data of ref. \cite{Astra2} show that this is indeed the case, and from their fit one can estimate $(\rho a)_c\approx 0.4-0.45$. For completeness, we note that, although our calculation is not quantitatively correct for $\rho a \approx 0.25-0.3$, the perturbative contact becomes negative at $(\rho a)_c \approx 0.38$.   

In a realistic experiment, the sTG gas is created under harmonic confinement \cite{Haller}. A qualitative picture of the trapped system can be inferred from the homogeneous Bose gas via the local density approximation (LDA) \cite{Stringari}. Within the LDA, we calculate the contact $\mathcal{I}$ for a harmonically trapped sTG gas ($W(x)=m\omega^2 x^2/2$), by making use of the virial theorem, Eq. (\ref{virial}), which is accessible with current experimental techniques \cite{Jin}. We plot it in Fig. \ref{fig-contact} as a function of $Na^2/a_{\mathrm{ho}}^2$, where $a_{\mathrm{ho}}=\sqrt{\hbar/m\omega}$. The appearance of a maximum for the contact and its subsequent depletion -- related to the gas-phase instability, as noted above for the homogeneous case -- would be clear experimental signatures of the sTG gas.

\section{Concluding remarks} 
We have shown that the Bose gas with hard-sphere interactions is equivalent to a many-boson system with momentum-dependent contact interparticle potentials. The resulting model is Fourier-transformable, and constitutes, for the first time, a simple momentum representation for the singular one-dimensional hard-sphere potential. As an important application, we have used our equivalent, soft-core model as a starting point to obtain the properties of the attractive and repulsive delta Bose gases in perturbation theory, and are in good agreement with the large-scale numerical simulations of ref. \cite{Astra2} in the limit of applicability of our perturbation theory. Universal Tan relations for the Lieb-Liniger gas are also derived and applied to estimate the contact for the super Tonks-Girardeau gas within our perturbation-theoretic approach.

The momentum-dependent zero-range interactions in this work have been carefully defined and are general. Therefore, they are applicable to other one-dimensional systems: we can use them to construct integrable approximations to non-integrable systems from their two-body phase shifts, and treat the difference between the original and model interactions as a weak perturbation.

\acknowledgments 
I am very grateful to Marvin D. Girardeau for many useful discussions during the final stages of his work in ref. \cite{AstraGirardeau}, and thank the authors of \cite{Astra2} for providing their Monte Carlo data for comparison. Useful correspondence with Murray T. Batchelor and Lukas F. Buchmann is gratefully acknowledged. The author was supported by a Villum Kann Rasmussen block scholarship.

\end{document}